\documentclass[a4paper,11pt]{article}
\usepackage{pos}

\title{Radiative energy loss of heavy quark through soft gluon emission in QGP}

\author*[a]{Taesoo Song}
\author[b]{Ilia Grishmanovskii}
\author[b,c]{Olga Soloveva}
\author[a,b,c]{Elena Bratkovskaya}

\affiliation[a]{GSI Helmholtzzentrum f\"{u}r Schwerionenforschung GmbH,\\ Planckstrasse 1, Darmstadt, Germany}

\affiliation[b]{Institut f\"ur Theoretische Physik, Johann Wolfgang Goethe-Universit\"at,\\ Max-von-Laue-Str.\ 1, Frankfurt am Main, Germany}

\affiliation[c]{Helmholtz Research Academy Hesse for FAIR (HFHF), GSI Helmholtz Center for Heavy Ion Physics,\\ Ruth-Moufang-Straße 1, Frankfurt am Main, Germany}

\emailAdd{T.Song@gsi.de}
\emailAdd{grishm@itp.uni-frankfurt.de}
\emailAdd{soloveva@itp.uni-frankfurt.de}
\emailAdd{E.Bratkovskaya@gsi.de}

\abstract{The Low's theorem is applied to the soft gluon emission from heavy quark scattering in quark-gluon plasma (QGP). The QGP is described by the dynamical quasi-particle model (DQPM) which reproduces the EoS from lQCD at finite temperature and chemical potential. We show that if the emitted gluon is soft and of long wavelength, the scattering amplitude can be factorized into the scattering part and the emission part and the Slavnov-Taylor identities are satisfied in the leading order. Imposing a proper upper limit on the emitted gluon energy, we obtain the scattering cross sections of charm quark as well as the transport coefficients (momentum drag and diffusion) in the QGP with and without gluon emission.}

\FullConference{
 26-31 March 2023 \\
 Aschaffenburg, Germany\\}

\begin{document}
\maketitle

\section{Introduction}

The energy loss of heavy flavor in heavy-ion collisions is one of important observables which reveal the properties of the quark-gluon plasma (QGP).
The energy loss takes place through elastic and inelastic scatterings~\cite{Uphoff:2012gb,Aichelin:2013mra,Berrehrah:2014kba}.
In this study we extend the soft photon approximation~\cite{Low:1958sn,Song:2018wvd,Song:2022ywu} to the soft gluon emission from the elastic scattering of heavy quark, and discuss its effect on the transport coefficients of heavy quark in QGP.

\section{Soft gluon approximation}

In the soft photon approximation~\cite{Low:1958sn,Song:2018wvd,Song:2022ywu} a low-energy photon is emitted from the external charged particles and the complicated inner structure of the scattering can be ignored.
Then the Feynman diagrams can be factorized into the elastic scattering and the gluon emission.

\begin{figure}[h]
\centerline{
\includegraphics[width=12. cm]{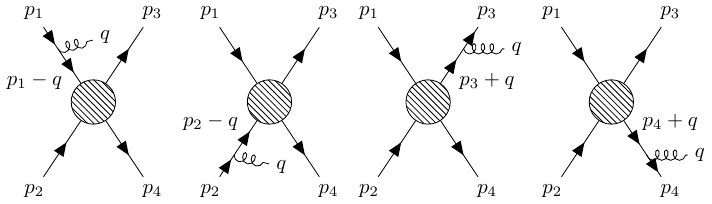}}
\vspace*{5mm}
\centerline{
\includegraphics[width=4.5 cm]{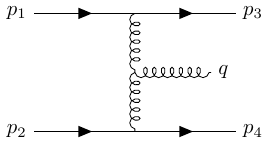}}
\caption{Feynman diagrams for $c+q\rightarrow c+q$ with a gluon emitted from external partons (upper) and from the internal exchanged gluon (lower).} 
\label{spa}
\end{figure}

Applying the same approximation to the soft gluon emission from $c+q\rightarrow c+q$ scattering, it is sufficient to calculate the upper diagrams in Fig.~\ref{spa}, of which transition amplitude is given by~\cite{Song:2022wil}
\begin{eqnarray}
M_{2q\rightarrow 2q+g}^{\mu, a, kl;ij}(p_1,p_2;p_3,p_4,q)=g
\bigg\{\frac{p_1^\mu}{p_1\cdot q}M_{2q\rightarrow 2q}^{kl;mj}T^a_{mi}\nonumber\\
+\frac{p_2^\mu}{p_2\cdot q}M_{2q\rightarrow 2q}^{kl;im}T^a_{mj}-\frac{p_3^\mu}{p_3\cdot q}T^a_{km}M_{2q\rightarrow 2q}^{ml;ij}-\frac{p_4^\mu}{p_4\cdot q}T^a_{lm}M_{2q\rightarrow 2q}^{km;ij}\bigg\},
\label{quark}
\end{eqnarray}
where $\mu,a$ are respectively the Lorentz and color indices of emitted gluon, $i,j$ and $k,l$ the color indices of the incoming and outgoing quarks and  $M_{2q\rightarrow 2q}^{kl;ij}$ the transition amplitude of  $c+q\rightarrow c+q$ scattering.
Considering the simplest color structure of $M_{2q\rightarrow 2q}^{kl;ij}$ is $T^b_{ki} T^b_{lj}$, one can prove that the transition amplitude of Eq.~(\ref{spa}) satisfies the Slavnov-Taylor identities~\cite{Song:2022wil}:
\begin{eqnarray}
q_\mu M_{2q\rightarrow 2q+g}^{\mu, a, kl;ij}(p_1,p_2;p_3,p_4,q)=0.
\label{conservation-q}
\end{eqnarray}


On the other hand, the lower diagram in Fig.~\ref{spa} 
is simplified in the limit $q\rightarrow 0$ into 
\begin{eqnarray}
M_5\approx -ig\varepsilon_\mu^{a*}(q) \frac{2(p_1-p_3)^\mu}{(p_1-p_3)^2+2(p_1-p_3)\cdot q} f^{abc}\bigg(M_{2q\rightarrow 2q}^{kl;ij}\bigg)^{bc}.
\label{exchange2}
\end{eqnarray}
Comparing Eqs.~(\ref{quark}) and (\ref{exchange2}), apart from $M_{2q\rightarrow 2q}$, the former is of the order of $1/q$ and the latter of the order of $1/p$ in the limit $q/p \ll 1$.
Therefore, the lower diagram is of higher order than the upper diagrams in Fig.~\ref{spa}, if the transition amplitude is expanded in term of $q/p$.

Similar calculations can be done for the soft gluon emission from $q+g\rightarrow q+g$, and the transition amplitude reads~\cite{Song:2022wil}.
\begin{eqnarray}
M_{q+g\rightarrow q+g+g}^{\mu,jbc;ia}(p_1,p_2;p_3,p_4,q)
=g\bigg\{\frac{p_1^\mu}{p_1\cdot q}M_{q+g\rightarrow q+g}^{jb;ma}T^c_{mi}
-\frac{p_3^\mu}{p_3\cdot q}T^c_{jm}M_{q+g\rightarrow q+g}^{mb;ia}\nonumber\\
+i\frac{p_2^\mu}{p_2\cdot q}f^{adc}M_{q+g\rightarrow q+g}^{jb;id}+i\frac{p_4^\mu}{p_4\cdot q}f^{bdc}M_{q+g\rightarrow q+g}^{jd;ia}\bigg\},
\label{spa2}
\end{eqnarray}
where $a, b$ and $c$ are respectively the colors of the incoming, outgoing and emitted gluons, and $i$ and $j$ the colors of the incoming and outgoing quarks.
It also satisfies the Slavnov-Taylor identities, considering that the color structure of $M_{q+g\rightarrow q+g}^{jb;ia}$ is given by $[T^a,T^b]_{ji}$ or $if^{abc}T_{ji}^c$~\cite{Song:2022wil}:
\begin{eqnarray}
q_\mu M_{q+g\rightarrow q+g+g}^{\mu, jbc;ia}(p_1,p_2;p_3,p_4,q)=0.
\label{conservation-g}
\end{eqnarray}

In QGP partons are dressed and interacting.
As a result, the parton self energy is a nonzero complex function whose real and imaginary parts are interpreted as $T,\mu_B-$dependent pole mass and spectral width, respectively.
The massive off-shell partons reproduce lattice EoS of QGP in the dynamical quasi-particle model (DQPM)~\cite{Moreau:2019vhw}.
Then the propagators of quark and gluon in Eqs.~(\ref{spa}) and (\ref{spa2}) are modified, respectively, into
\begin{eqnarray}
G_{ij}(p)&=&i\frac{\not{p}+m_q(T,\mu)}{p^2-m_q^2(T,\mu)+i|p_0|\Gamma_q(T,\mu)}\delta_{ij},\nonumber\\
G^{\mu\nu,ab}(p)&=&\frac{-ig^{\mu\nu}}{p^2-m_g^2(T,\mu)+i|p_0| \Gamma_g(T,\mu)}\delta_{ab},
\end{eqnarray} 
where $m_{q(g)}$ and $\Gamma_{q(g)}$ are the pole mass and spectral width of quark(gluon).

Assuming the center-of-mass frame of $p_3+p_4$ is similar to that of $p_1+p_2$, the differential cross section is approximated as~\cite{Song:2022wil}
\begin{eqnarray}
\frac{d\sigma_{2\rightarrow 3}}{d\cos\theta}
\approx \frac{d\sigma_{2\rightarrow 2}}{d\cos\theta}\int_{m_g}^{E_{max}} \frac{dE_g}{2(2\pi)^3} \sqrt{E_g^2-m_g^2} \int d\cos\theta' d\phi'  |\epsilon\cdot J|^2\frac{|{\bf p}_3|\sqrt{s}}{p_f\sqrt{s_2}},
\label{approx1}
\end{eqnarray}
where $\theta$ is the scattering angle, ($\theta'$, $\phi'$) the solid angle of emitted gluon, $m_g$ the thermal gluon mass, $s_2=(p_3+p_4)^2$  and 
\begin{eqnarray}
\overline{|M_{2\rightarrow 3}}|^2 \equiv |\epsilon\cdot J|^2~ \overline{|M_{2\rightarrow 2}}|^2
=32\pi s|\epsilon\cdot J|^2\frac{p_i}{p_f}\frac{d\sigma_{2\rightarrow 2}}{d\cos\theta}
\label{def1}
\end{eqnarray}
with $p_i$ and $p_f$ being respectively the initial and final momenta of 2-to-2 scattering in the center-of-mass frame.
The maximum energy of soft gluon, $E_{max}$ in Eq.~(\ref{approx1}),  is kinetically given by $E_{max}^{kin.}=\sqrt{m_g^2+q_{max}^2}$ with $q_{max}^2=\{s-(m_3+m_4+m_g)^2\}\{s-(m_3+m_4-m_g)^2\}/(4s)$.
In the soft gluon approximation (SGA), however, we have neglected gluon emission from the interaction region, which means that the wavelength of the emitted gluon is assumed larger than the scattering scale that is roughly $1/\sqrt{-t}$  with $t=(p_1-p_3)^2$. Therefore, more reasonable upper limit for SGA will be $E_{max}^{real.}={\rm min}\bigg[E_{max}^{kin.},~\sqrt{-t}\bigg]$.

\begin{figure}[h]
\centerline{
\includegraphics[width=8.3 cm]{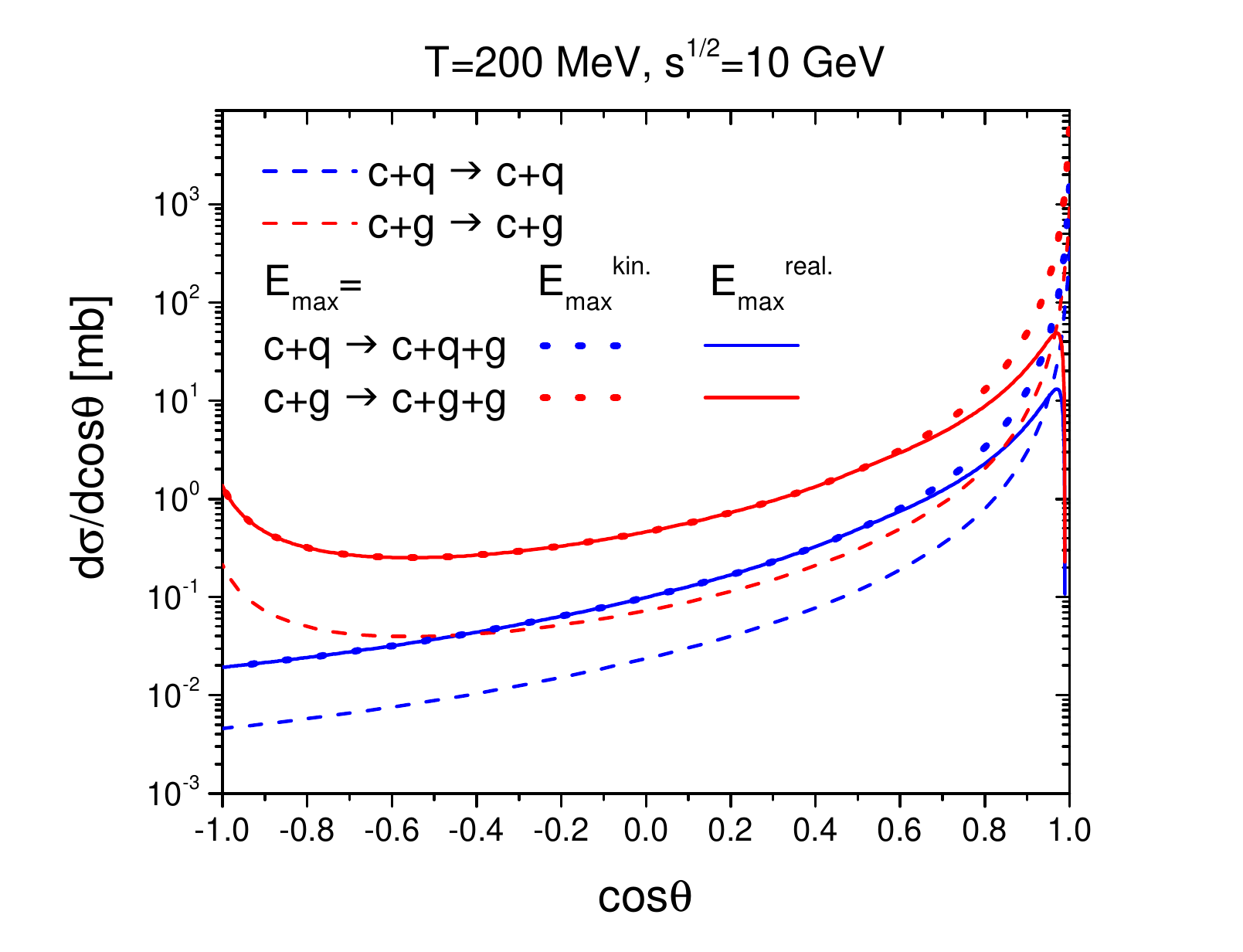}
\includegraphics[width=8.3 cm]{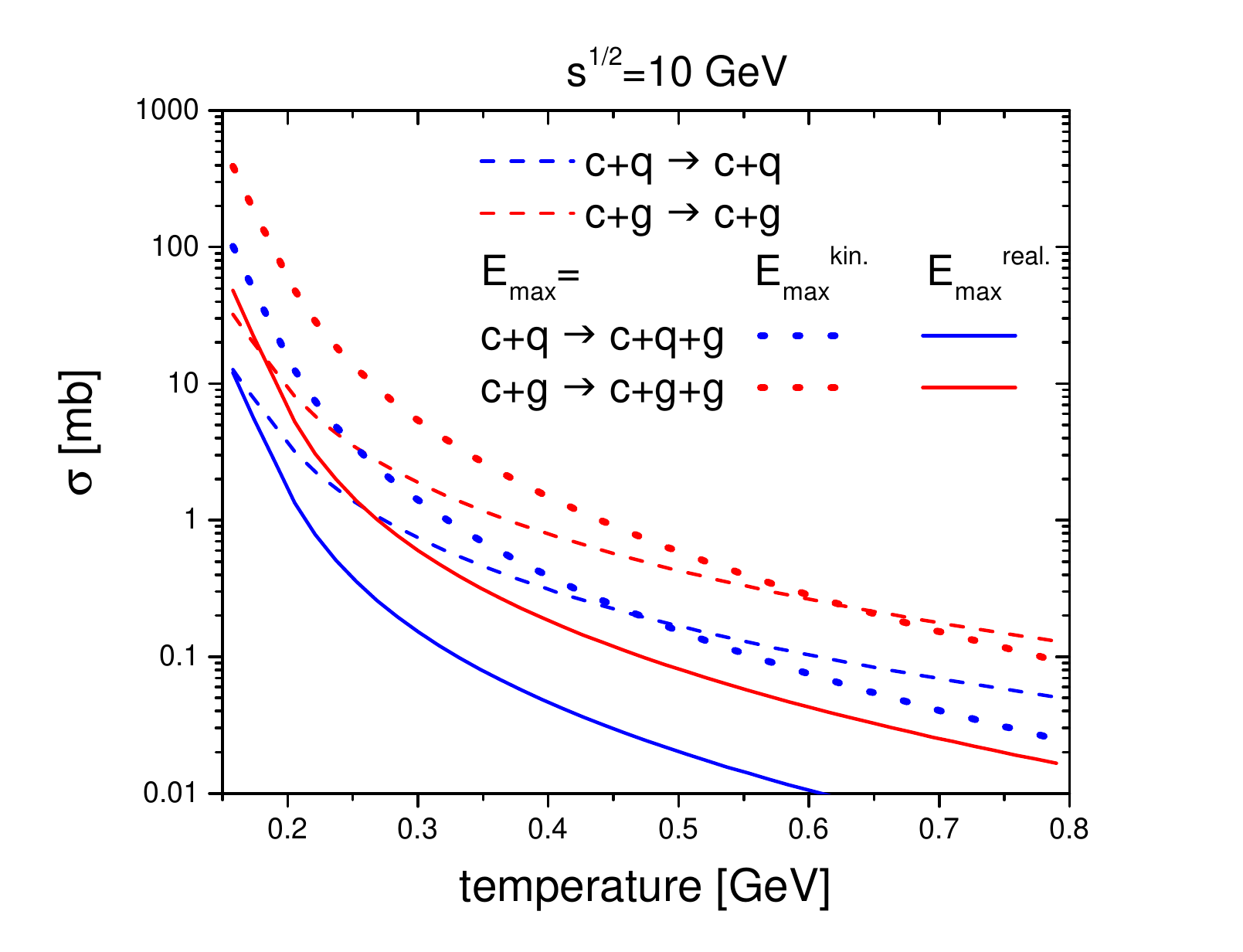}}
\caption{Differential (left) cross sections of $c+q\rightarrow c+q+g$ and $c+g\rightarrow c+g+g$ in the soft gluon approximation with the kinetic and the realistic energy limits, which are respectively shown as dotted and solid lines, in comparison with the elastic scattering cross sections, which are dashed lines, at $\sqrt{s}=$10 GeV and $T=$ 200 MeV and (right) their integrated cross sections as a function of temperature at $\sqrt{s}=$10 GeV.}
\label{sigma}
\end{figure}

The left panel of Fig.~\ref{sigma} displays the differential cross sections for $c+q(g)\rightarrow c+q(g)$ and $c+q(g)\rightarrow c+q(g)+g$ with the kinematic and realistic upper limits of gluon energy as a function of the scattering angle of charm quark at $\sqrt{s}=$10 GeV and $T=$ 200 MeV.
Charm quark mass is taken to be 1.5 GeV, and the masses of light quark and gluon the pole masses in the DQPM.
The radiative differential cross sections with the two different upper limits are almost same except near $\cos\theta=1$, where $t$ is small and thus the realistic upper limit of the soft gluon energy is low.
In the right panel the integrated cross sections are shown as a function of temperature.
The temperature dependence of 2-to-3 scattering cross sections is stronger than that of 2-to-2 scattering cross sections, because the former is proportional to $\alpha_s^3$ while the latter to $\alpha_s^2$, and the $\alpha_s$ rapidly increases near $T_c$ in the DQPM.
We note that the above results agree with those from the full calculations without introducing the SGA at low momentum of emitted gluon~\cite{ilia}.



\section{Transport coefficients}

Heavy quark loses energy-momentum traversing a QGP produced in relativistic heavy-ion collisions.
The most suitable quantities to describe the energy-momentum loss are the momentum drag coefficient ($A$) and the transverse momentum transfer squared per unit length $\hat{q}$ which are defined as
\begin{eqnarray}
A(p)&=& -\frac{d\langle \Delta p_L \rangle}{dt}=\langle\langle(p-p^\prime)_z\rangle\rangle,\label{A}\\
\hat{q}(p)&=&\frac{d\langle(\Delta p_T)^2\rangle}{d z}=\frac{E}{p_L}\langle\langle p_x^{\prime 2}+p_y^{\prime 2}\rangle\rangle,
\label{qhat}
\end{eqnarray}
where $p_L$ and $p_T$ are respectively the longitudinal and transverse momentum of charm quark with $p$ and $p^\prime$ being respectively initial and final momenta, and the double bracket implies
\begin{eqnarray}
\langle\langle O^* \rangle\rangle \equiv \sum_{i=q,\bar{q},g}\int dm_i dm_f dm_g A_i(m_i)A_i(m_f)A_g(m_g)
\int \frac{d^3k}{(2\pi)^3 }f_i(k)~O^*~ v_{ic}\sigma_{ic},
\label{def2}
\end{eqnarray}
where $m_i$, $m_f$ and $m_g$ are respectively the incoming and outgoing parton masses and the emitted gluon mass, $A_i$, $A_f$ and $A_g$ their spectral functions in the DQPM,  $f_i(k)$ is a distribution function of parton $i$, and $v_{ic}$ and $\sigma_{ic}$ are the relative velocity and the scattering cross section of the charm quark and parton $i$, respectively.

\begin{figure}[h]
\centerline{
\includegraphics[width=8.3 cm]{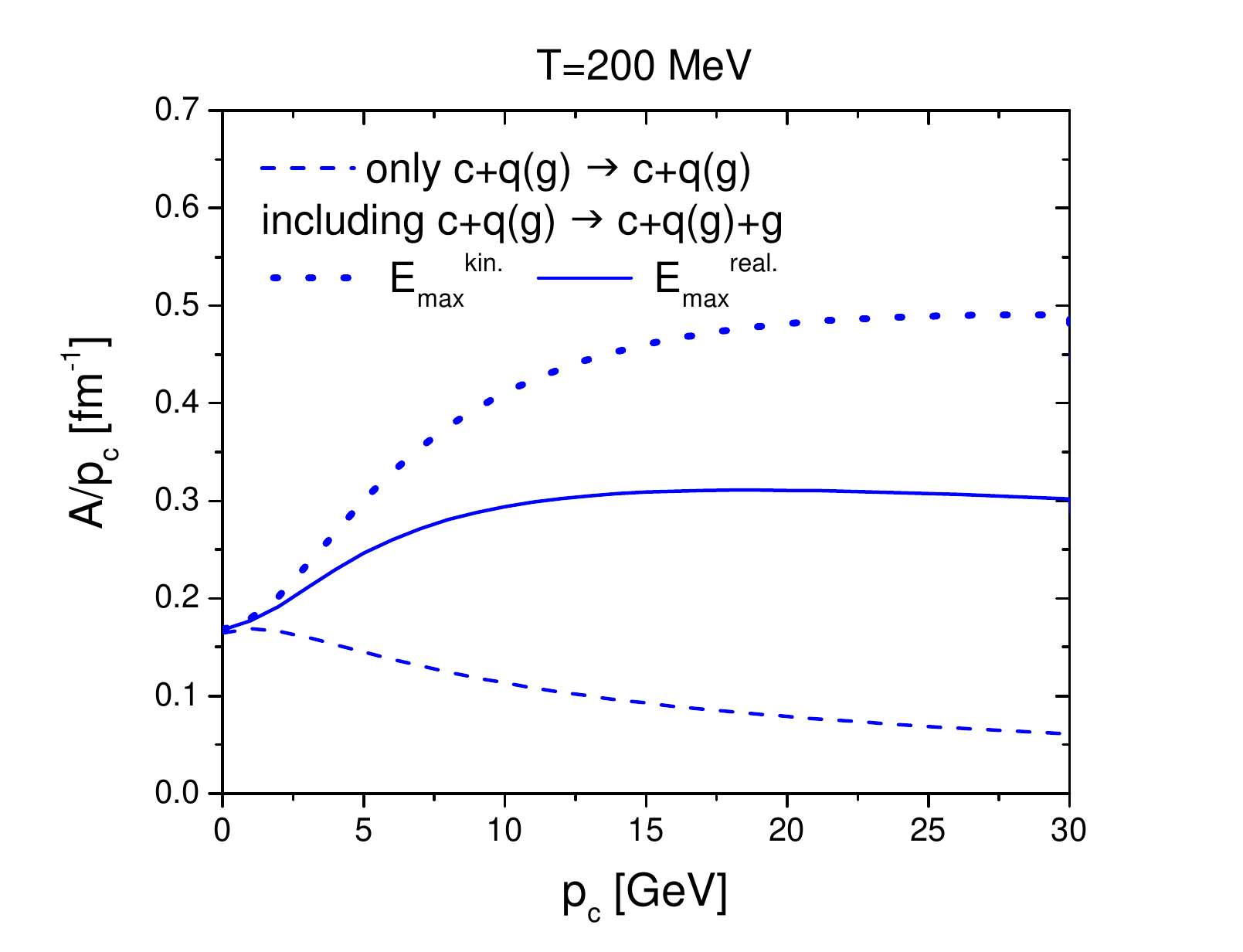}
\includegraphics[width=8.3 cm]{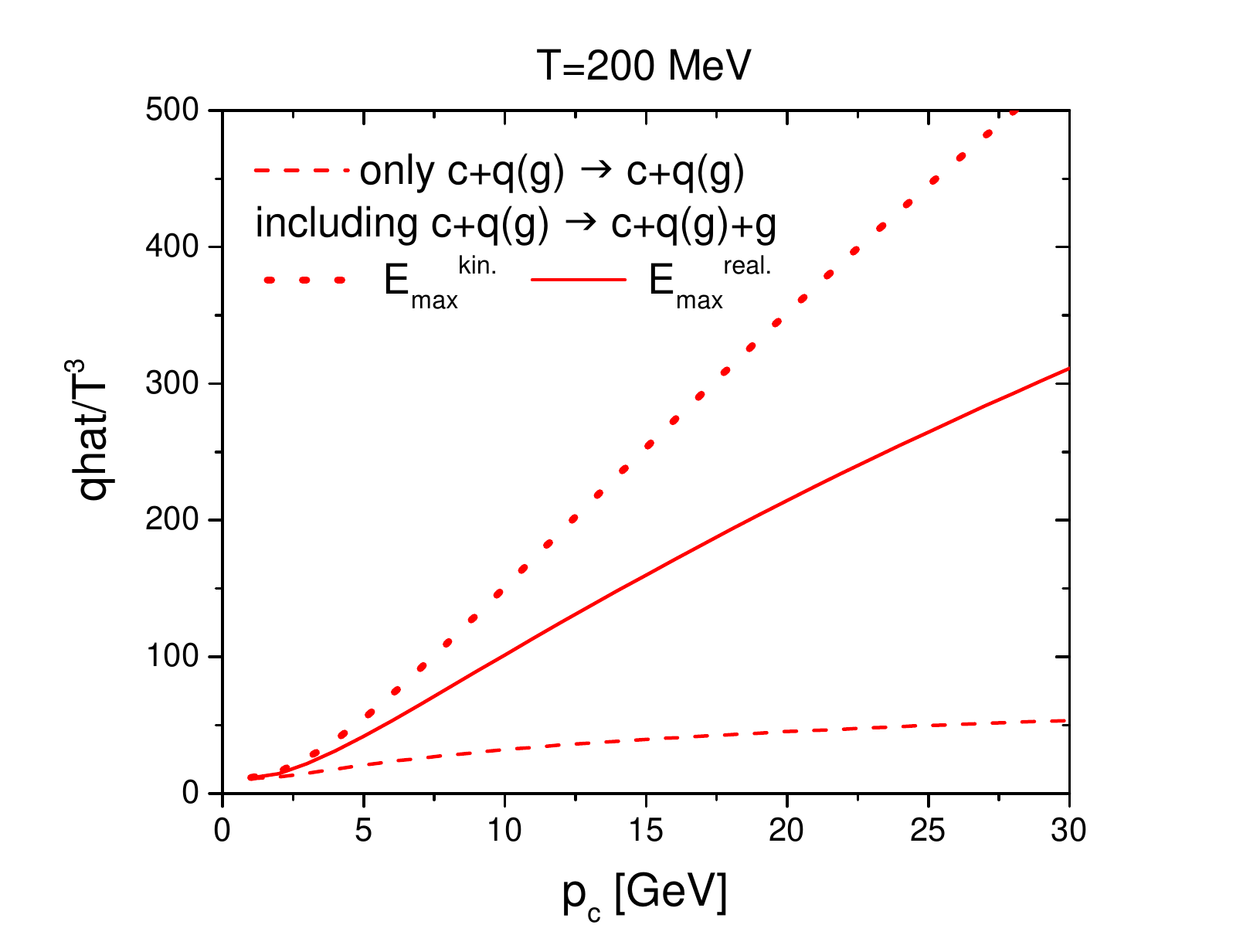}}
\caption{Momentum drag coefficient $\eta_D=A/p_c$ (left) and $\hat{q}/T^3$ (right) of charm quark as a function of charm quark momentum at $T=$ 200 MeV, excluding and including the radiative corrections with the kinetic and the realistic upper limits of emitted gluon energy.}
\label{transport}
\end{figure}

In Fig.~\ref{transport} we present the drag coefficients $\eta_D=A/p_c$ (left) and $\hat{q}/T^3$ (right) of the charm quark without and with soft gluon emission for the kinematic and realistic upper limits of soft gluon energy at $T=200$ MeV.
The radiative corrections hardly change the transport coefficients at low momentum, because the scattering energy of a slow charm quark with a thermal parton is not large enough to produce an additional massive gluon.
As a result, the radiative scattering barely contributes to the spatial diffusion coefficient $D_s =\lim_{p_c\to 0} T/(M_c\eta_D)$ of heavy quark~\cite{Song:2022wil}, which is already well reproduced within the DQPM~\cite{Berrehrah:2014kba,Song:2016lfv}.

The radiative energy loss, however, enhances $A$ and $\hat{q}$ of the charm quark at large momentum.
It may 
influence the $R_{\rm AA}$ of $D$ mesons which has been well described only with elastic scattering in heavy-ion collisions~\cite{Song:2015sfa,Song:2015ykw}.
According to our recent study~\cite{Grishmanovskii:2022tpb} the $\alpha_s (T)$ extracted from the lattice EoS overestimates jet quenching.
More realistic strong coupling at large momentum will be smaller than $\alpha_s (T)$, because an energetic parton is far off thermal equilibrium, and it will cure the potential problem raised from the enhancement of transport coefficients in Fig.~\ref{transport}.

\section{Conclusion}

We have studied the radiative energy loss of heavy quark by using the soft gluon approximation which is motivated by the soft photon approximation in QED.
It has been shown that for the pQCD case this approximation satisfies the Slavnov-Taylor identities, which guarantees the gauge invariance, and the results agree with those from the full calculations at the low energy of emitted gluon.
Extending it to the energy loss in non-perturbative strongly interacting QGP within the DQPM, we have found that the radiative energy loss little affects the spatial diffusion coefficient $D_s$ of heavy quark, while it greatly enhances the transport coefficients at large momentum of heavy quark, which can be suppressed by adopting a more realistic $\alpha_s$ for the energetic scattering of partons.   

\vspace{3mm}
The authors acknowledge  support by the Deutsche Forschungsgemeinschaft (DFG, German Research Foundation) through the grant CRC-TR 211 "Strong-interaction matter under extreme conditions" - project number 315477589 - TRR 211. I.G. also acknowledges support from the "Helmholtz Graduate School for Heavy Ion research". The computational resources have been provided by the Goethe-HLR Center for Scientific Computing.

\end{document}